# Recommendation Algorithm Based on Recommendation Sessions

Michał MALINOWSKI

Military University of Technology / Faculty of Cybernetics / Institute of Computer and Information Systems
Warsaw, Poland, michal.malinowski@wat.edu.pl

## Abstract

The enormous development of the Internet, both in the geographical scale and in the area of using its possibilities in everyday life, determines the creation and collection of huge amounts of data. Due to the scale, it is not possible to analyse them using traditional methods, therefore it makes a necessary to use modern methods and techniques. Such methods are provided, among others, by the area of recommendations.

The aim of this study is to present a new algorithm in the area of recommendation systems, the algorithm based on data from various sets of information, both static (categories of objects, features of objects) and dynamic (user behaviour).

**Keywords**: Recommendation System and Algorithm, Graphs and Networks, E-Commerce.

## Introduction

Recommendation Systems (RS) are currently ubiquitous in many areas of life. They become indispensable tools to help users to search for appropriate information in the huge amount of data that surrounds them. Such systems influence the decisions not only of decision-makers, but also of ordinary citizens in their daily lives. It is estimated that in 2020 the level of about 40 trillion gigabytes of data stored in social media databases, electronic mail systems (e-mails) and indexed websites will be reached.

## Outline of algorithms of recommendation systems

The currently existing recommendation systems could be divided into three main techniques:

- **Content Based** (methods of filtering based on content) – such systems are constructed on the assumption that objects with similar features will receive similar ratings from the same user;
- **Collaborative Filtering - CF** (methods of filtering based on neighbourhood) - such systems are based on the idea, that if two users have made similar decisions in the past, their preferences will also converge in the future. This method can be divided into the following:
    - **Model Based;**
    - **Memory Based:**
        - **User Based;**
        - **Item Based.**
    - **Hybrid Collaborative Filtering.**
- **Hybrid Filtering** (hybrid algorithms) – systems using the rules of the two previous techniques.

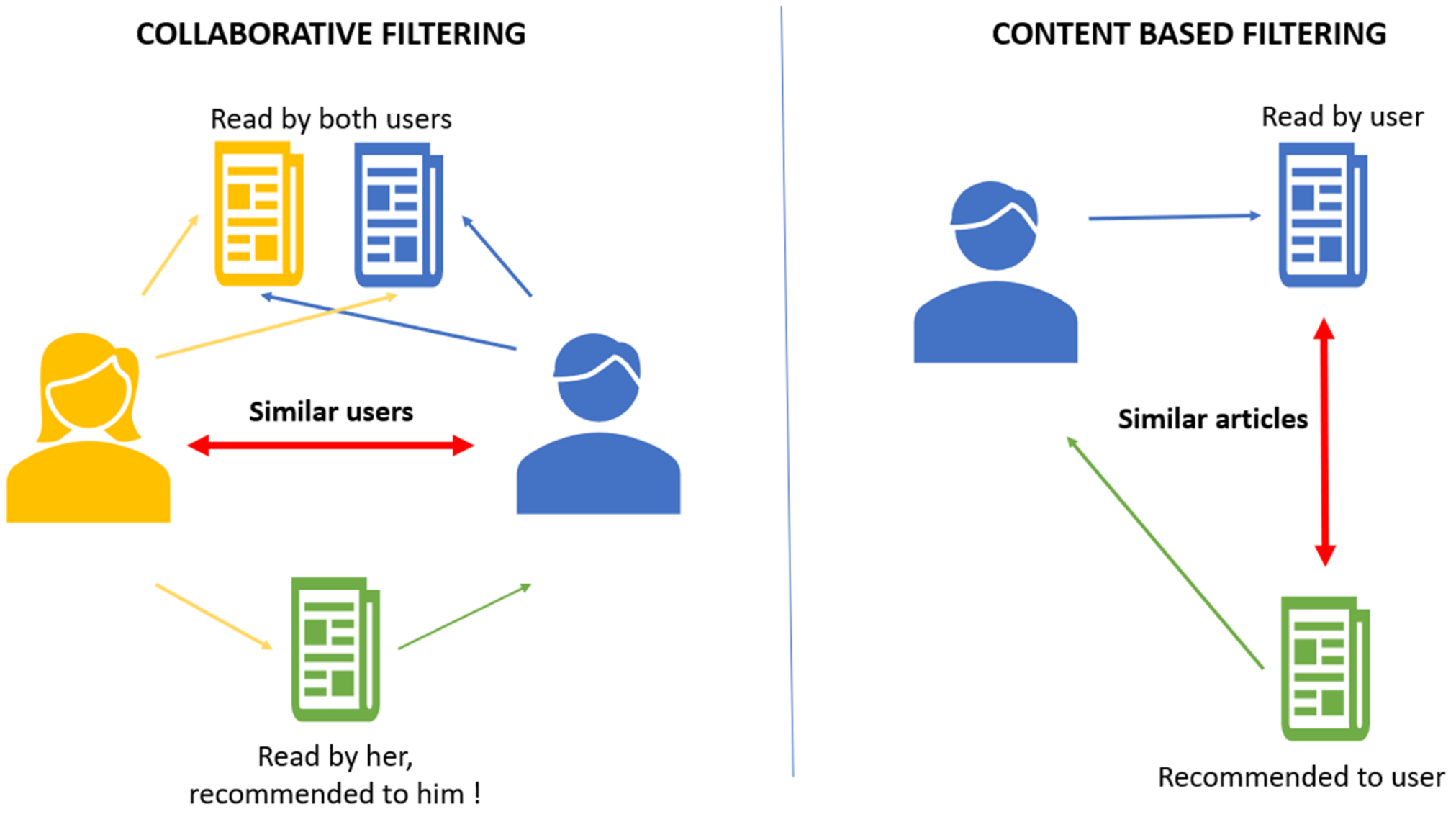

**Fig. 1: The idea of operation of the systems based on Content Based Filtering and Collaborative Filtering Methods.**

Table 1. provides an overview, advantages and disadvantages of the algorithms used within the above mentioned types of methods [4].

**Table 1: Comparison of recommendation techniques**

| | |
|---|---|
| Technique | Content Based |
| Algorithms | Content Based algorithm using Hidden Markov Model |
| Advantages | Disadvantages |
| No scarcity and cold start problem<br>It ensures privacy | Requires rich description of items<br>Requires organized user profile<br>Content overspecialization is problem |
| Technique | Collaborative Filtering (Memory Based) |
| Algorithms | User Based CF<br>Item Based CF |
| Advantages | Disadvantages |
| Easy to implement<br>Data addition is simple<br>Need not consider content<br>Good scalability | Depends on explicit feedback<br>Cold start problem<br>Sparsity problem<br>Limited scalability for large dataset |
| Technique | Collaborative Filtering (Model Based) |
| Algorithms | Slope-one CF<br>Matrix factorization |
| Advantages | Disadvantages |
| Improving prediction performance<br>Improving scalability and sparsity problem | Model is expensive<br>Loss of information in matrix factorization |
| Technique | Hybrid Collaborative Filtering |
| Algorithms | Combination Model Based and Memory Based |
| Advantages | Disadvantages |
| Overcoming limitations of sparsity<br>Improving prediction performance | Increased complexity<br>Complicated for implementation |
| Technique | Hybrid Filtering |
| Algorithms | Combination Content Based and Collaborative Filtering |
| Advantages | Disadvantages |
| Collaborative and content based methods are complementary in their strengths and weaknesses<br><br>Overcoming scarcity and cold start problem | Difficult for implementation |

In algorithms of recommender techniques based on Content Based Filtering, user profiles include features and items chosen by the user in the past. Consequently, those algorithms have two main weaknesses: they only recommend items similar to those chosen in the past, and also, the system must extract meaningful and useful features from the description of items.

Collaborative Filtering's recommendation systems are based on the received information from users who shared similar interests and preferences in the past. In this method, algorithms use the feedbacks received from other similar users to give recommendations. The main advantage of this method over the previous one is that, in this technique, the community of users gives opinions about certain items, and also provide feedbacks and ratings. As a result, there is a chance for quite new items to be recommended to the user.

Considering the nature of this method, it provides the system will return weak and non-transparent predictions, when the number of users will be too low (Cold Start)

The algorithm described in the study can be classified as the hybrid algorithm type, because it can make recommendations based on both the similarity of objects' features (Content Based) and the similarity of users' behaviours (Collaborative Filtering - CF). Precise classification can be made only on the basis of the input data provided in the form of the graph **G,** which is described below.

The concept of a recommendation session may be reflected, for example, by the relationship between purchased orders and order items or between a product category and products. In the area of mathematical apparatus it is based on graph and network theory, and can be described on the basis of a bipartite directed graph **G** called the graph of the recommendation session such that:

$$G = \langle N, E \rangle \quad (1)$$

where:

$N = J \cup O$ – the collection of nodes

$E \subset J \times O$ – the collection of edges (directed edges)

$O$ – the collection of objects

**The object** ($o \in O$) may be:

- products in an Internet shop;
- a video in a video-on-demand service;
- an employee in an employment-related service;
- press article in the news service;
- a person on a social network application.

$J$ – collection of kernels

**The kernel** ($j \in J$) may be:

- a products' category – one of the sub-collections of products having common features;
- the purchase order – the result of the customer's actions in the shop, which is finalised with the purchase;
- a wish list – a sub-collection of the shop's products related to the customer, which such sub-collection results from the customer's future shopping preferences;
- an expert – a sub-collection of products indicated by the discipline specialist;
- a website visit ID – a unique key assigned to a user's visits to the website. A visit consists of a sequence of viewed/visited pages of an online store;

- a person – IT structure identifying and describing the user in the IT system.

Due to the adopted and further discussed interpretation of the model and its components, the following limitations are assumed to be met:

$J \cap O = \emptyset$ – no kernel shall be a facility and no facility shall be a kernel

$(\forall j \in J)(\exists e \in E)(e = (j, o))$ – each kernel must be associated with at least one object

$(\forall o \in O)(\exists e \in E)(e = (j, o))$ – each object must be associated with at least one kernel.

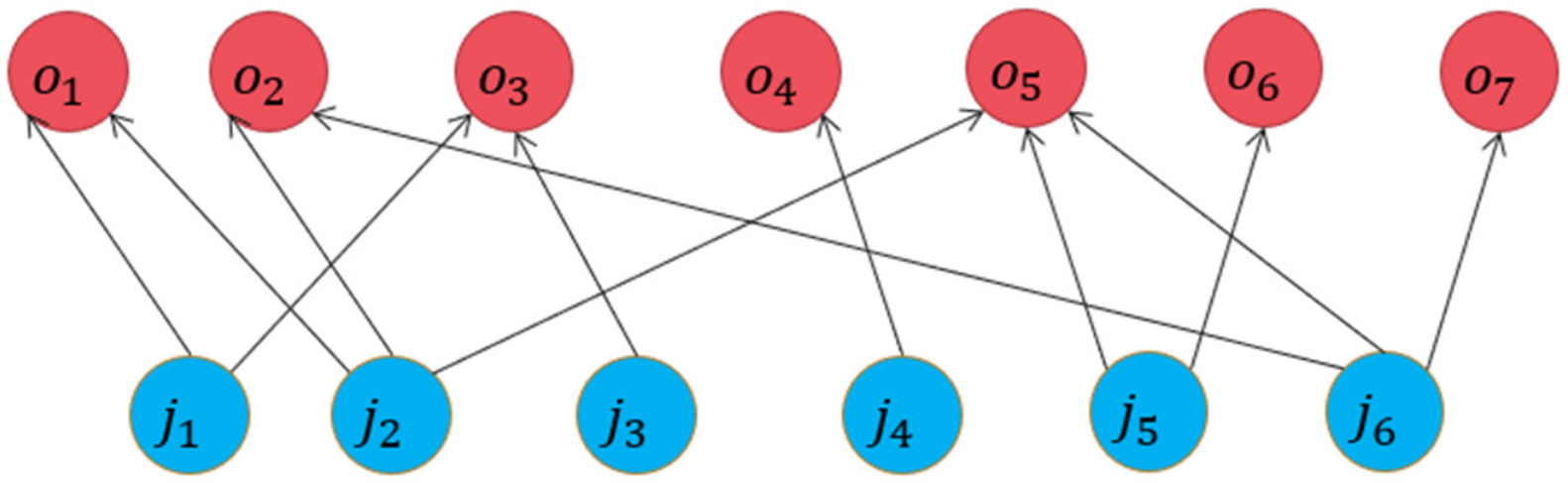

Fig. 2. The recommendation session's graph

A single session **S** can be presented as a sub-graph of the graph **G** such that:

$S =< N', E' >$ (2)

where:

$N' = j \cup O'$ – a collection of session's nodes

$E' \subset E$ – collection of session edges

$O' \subset O$ – a collection of session's objects (related to a kernel)

$j \in J$ – a session's kernel

$(\forall o \in O')(\exists e \in E)(e = (j, o))$ – each object of a session is associated with a kernel

$(\nexists o \in O \backslash O')(\exists e \in E)(e = (j, o))$ – there is no kernel-related object that is not a part of the session.

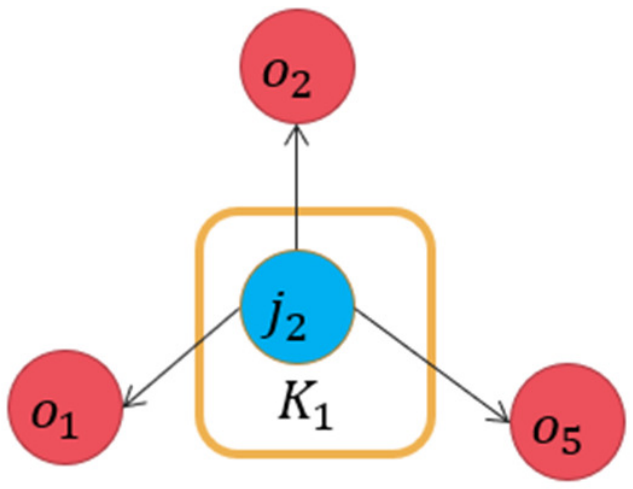

Fig. 3. Single recommendation session model

The session is made up of one kernel only and of all edges and objects related to the kernel.

**The sessions** ($S$) may be:

- the products' category and the products themselves;
- the purchase order and the order's items;
- the customer's wish list with the elements;
- an expert with his/her assessments;
- an identifier (ID) of the website visited, along with the pages visited;
- a person with his/her colleagues.

In addition, due to its physical properties and similarities, the session's kernels can be grouped into classes **K.** The kernels of the same type are grouped into the classes, e.g.: products' categories, purchase orders, etc.

Where:

$K \subset J$- the class is the sub-collection of the kernels' collection

**The class** ($K \subset J$) of the session's kernels may be:

- the categories of the products;
- the purchase orders;

- the customers' wish lists;
- experts;
- identifiers (IDs) of the website visits;
- persons.

In addition:

$(\forall i \neq j)(K_i \cap K_j = \emptyset)$ – no kernel is present in two or more classes

$\bigcup_{i=1}^{|K|} K_i = J$ – each kernel belongs to a class

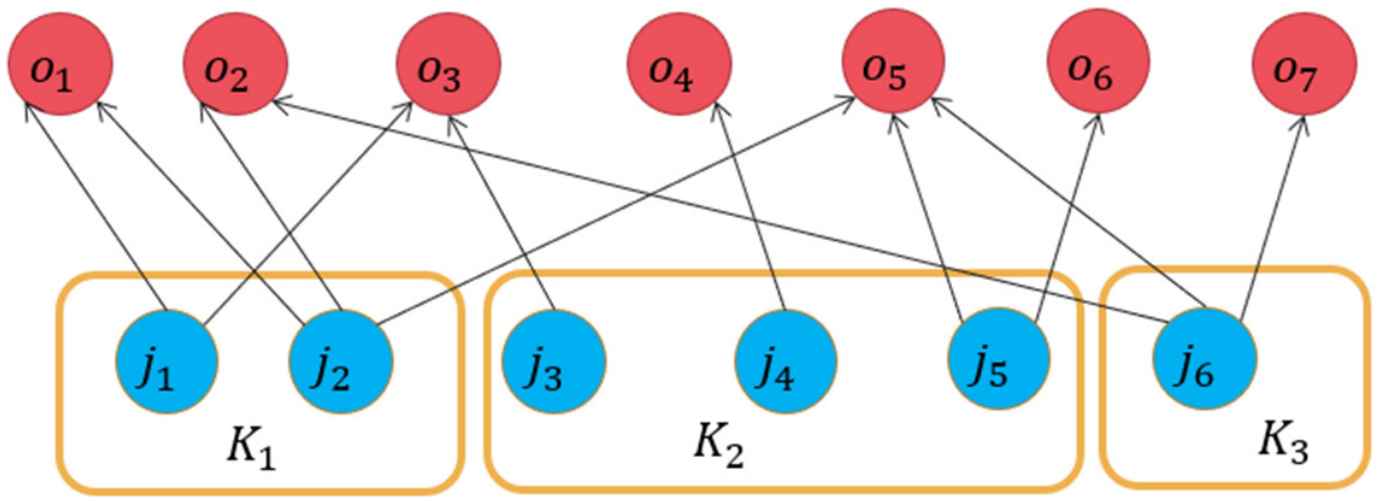

Fig. 4. The classes of the kernels

## Algorithm of the Recommendation Session

Let $\boldsymbol{C}$ be a collection of users using the recommendation system and $\boldsymbol{O}$ a collection of all possible objects that can be recommended.

In addition, let $\boldsymbol{u}$ be a utility function measuring the utility of object $\boldsymbol{o} \in \boldsymbol{O}$ for user $\boldsymbol{c} \in \boldsymbol{C}$. This is $\boldsymbol{u}: \boldsymbol{CxO} \rightarrow \boldsymbol{W}$, where $\boldsymbol{W}$ is an ordered collection (e.g. non-negative integers)

Then, for each user $\boldsymbol{c}$ belonging to $\boldsymbol{C}$, a sub-collection $\boldsymbol{R_c}$ belonging to $\boldsymbol{O}$ is selected, which maximises the usability for the user. This means that for the arranged $\boldsymbol{c} \in \boldsymbol{C}$ we determine the collection of the recommended objects as follows:

$$R_c = \{r \in O: u(c,r) = \max_{o \in O} u(c,o)\} \qquad (3)$$

In a general case, the task of the recommendation algorithms is to find a sub-collection $\boldsymbol{R_c}$ called the collection of recommended objects for user **c**.

In a specific case, a recommendation is made to the selected object $\boldsymbol{m}$ without directly considering the user $\boldsymbol{c}$, for whom the recommendation is made. Therefore, condition (3), taking into account a change in the function of the recommendation such that $\boldsymbol{u}: \boldsymbol{OxO} \rightarrow \boldsymbol{W}$, can be presented as follows:

– firstly, the collection of recommendations for user $\boldsymbol{R_c}$ is replaced by $\boldsymbol{R_m}$, i.e. the collection of recommendations for the object;

– secondly, for the arranged $\boldsymbol{m} \in \boldsymbol{O}$, we determine the collection of recommended objects as follows:

$$R_m = \{r \in O: u(m,r) = \max_{o \in O} u(m,o)\} \qquad (4)$$

Considering user $\boldsymbol{c}$ consists in the fact that it is him who, consciously directing his behaviour, selects an object $\boldsymbol{m}$ from the $\boldsymbol{O}$ collection, against which recommendation $\boldsymbol{R_m}$ shall be built up.

ARS (Algorithm of the Recommendation Sessions) based on this particular case.

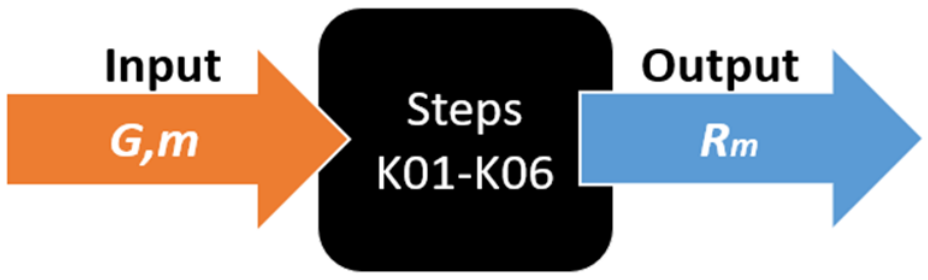

**Fig. 5: Architecture of the Algorithm of the Recommendations Session (ARS)**

<u>Input data:</u>

**G** – the session graph

**m –** object (graph node) to which the recommendations are to be linked

<u>Output data:</u>

**R<sub></sub>**

**$R_m$** – vector of recommendation for the object **m** (the lower the position in the vector, the better the recommendation for the object **m**)

<u>Steps:</u>

***K01**:* Presentation of input data

Example:

**G** $=< N, E >$

**m** = $o_3$

where:

$N = J \cup O$

$O = \{o_n : n = \overline{1,9}\}$

$J = \{j_n : n = \overline{1,6}\}$

$E = \{< j_1, o_1 >, < j_1, o_2 >, < j_1, o_5 >, < j_2, o_2 >, < j_2, o_3 >, < j_2, o_4 >, \ldots, < j_6, o_1 >, < j_6, o_6 >\}$

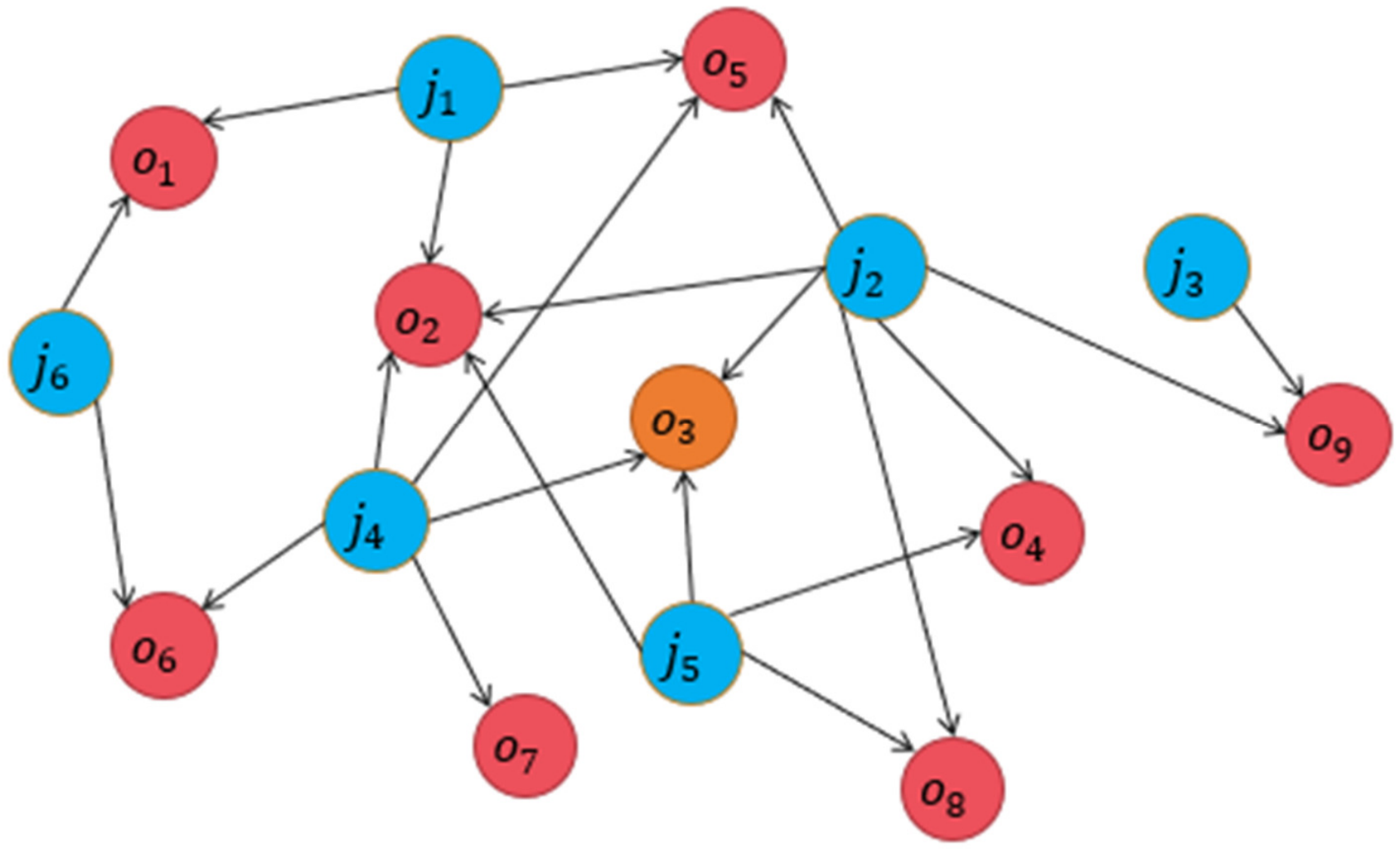

**Fig. 6: The graph of the input session G with the input object (node) marked m**

***K02:*** The design of the sub-graph **G'$_m$** of the graph **G** consisting of node **m** and nodes adjacent to the node **m** and edges between them and node **m.**

Example:

**G'$_m$** $=< N', E' >$

**m** = $o_3$

where:

$N' = O' \cup J'$

$O' = \{o_3\}$

$J' = \{j_2, j_4, j_5\}$

$E' = \{< j_2, o_1 >, < j_4, o_1 >, < j_5, o_1 >\}$

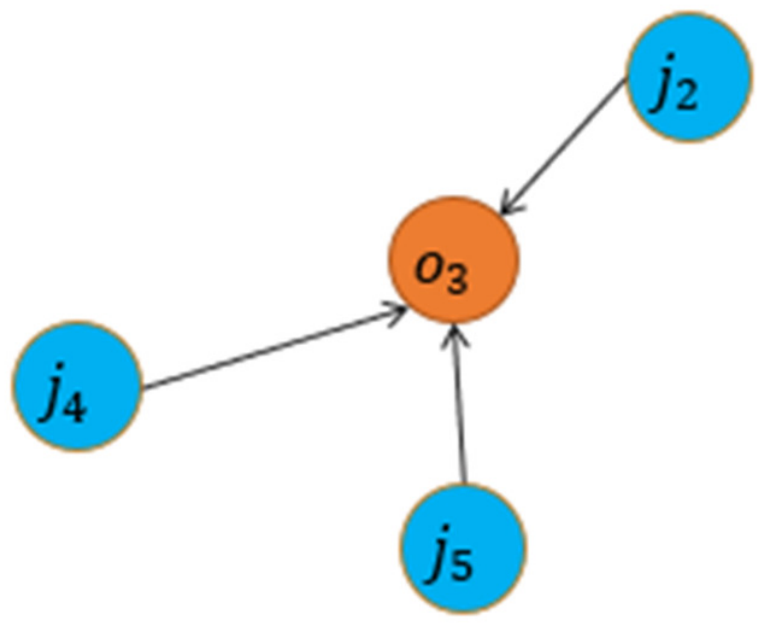

**Fig. 7: The designed sub-graph G'$_m$ of the graph G**

***K03****:* The construction of the sub-graph **G''$_m$** of the graph **G** composed of the graph **G'$_m$** and the nodes adjacent to the nodes of the graph **G'$_m$** and edges between them and the nodes of the graph **G'$_m$**

Example:

**G''$_m$**$=< N'', E'' >$

**m** $= o_3$

where:

$N'' = O'' \cup J''$

$O'' = \{o_3, o_2, o_4, o_5, o_6, o_7, o_8, o_9\}$

$J'' = \{j_2, j_4, j_5\}$

$E'' = \{< j_2, o_1 >, < j_4, o_1 >, < j_5, o_1 >, < j_2, o_4 >, < j_2, o_5 >, < j_2, o_9 >, < j_4, o_2 >, < j_4, o_6 >, < j_4, o_7 >, < j_5, o_2 >, < j_5, o_8 >\}$

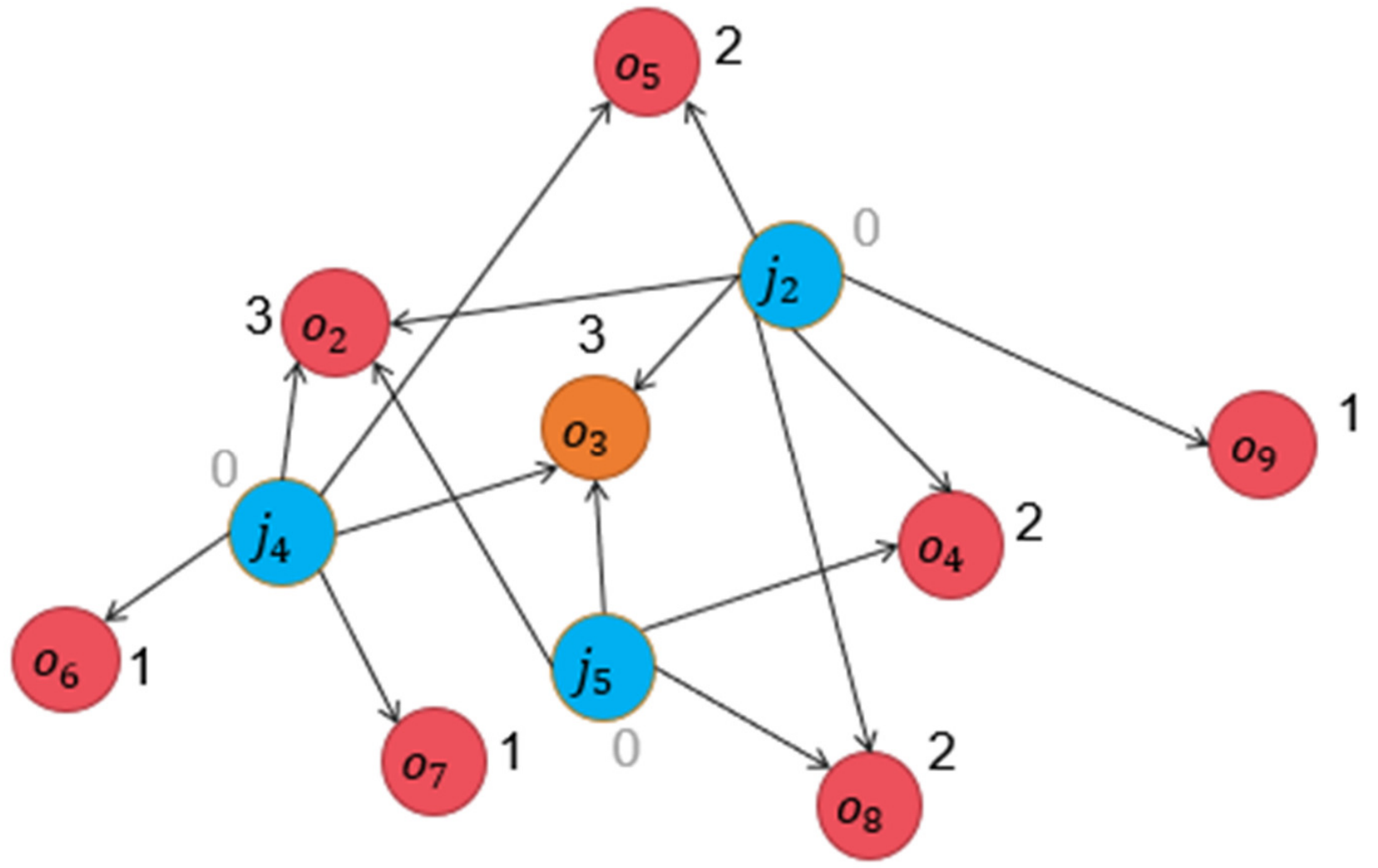

**Fig. 8: The constructed sub-graph G''$_m$ of the graph G with estimated incoming steps of the nodes**

***K04:*** Estimation of the incoming steps for each node that is the object of sub-graph **G''$_m$**

Example:

$deg_{in}(o_3) = 3$

$deg_{in}(o_2) = 3$

$deg_{in}(o_4) = 2$

$deg_{in}(o_5) = 2$

$deg_{in}(o_6) = 1$
$deg_{in}(o_7) = 1$
$deg_{in}(o_8) = 2$
$deg_{in}(o_9) = 1$

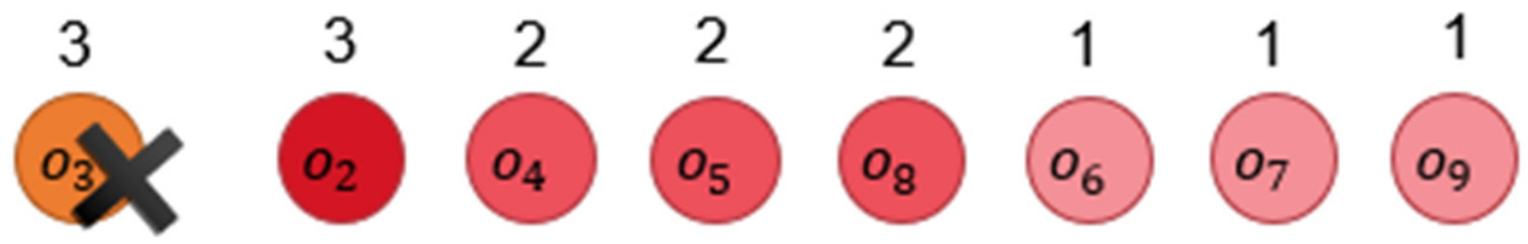

**Fig. 9: Sorted objects (nodes) with the removed object m**

## Application of the algorithm

The algorithm, as the part of the scientific research, was used to build recommendations for a real existing e-commerce solution based on real data. The precise data were downloaded from the relational database (MySQL) of the online store's IT system AM76 (https://am76.pl) for the time period from the 15th to the 20th of December 2019.

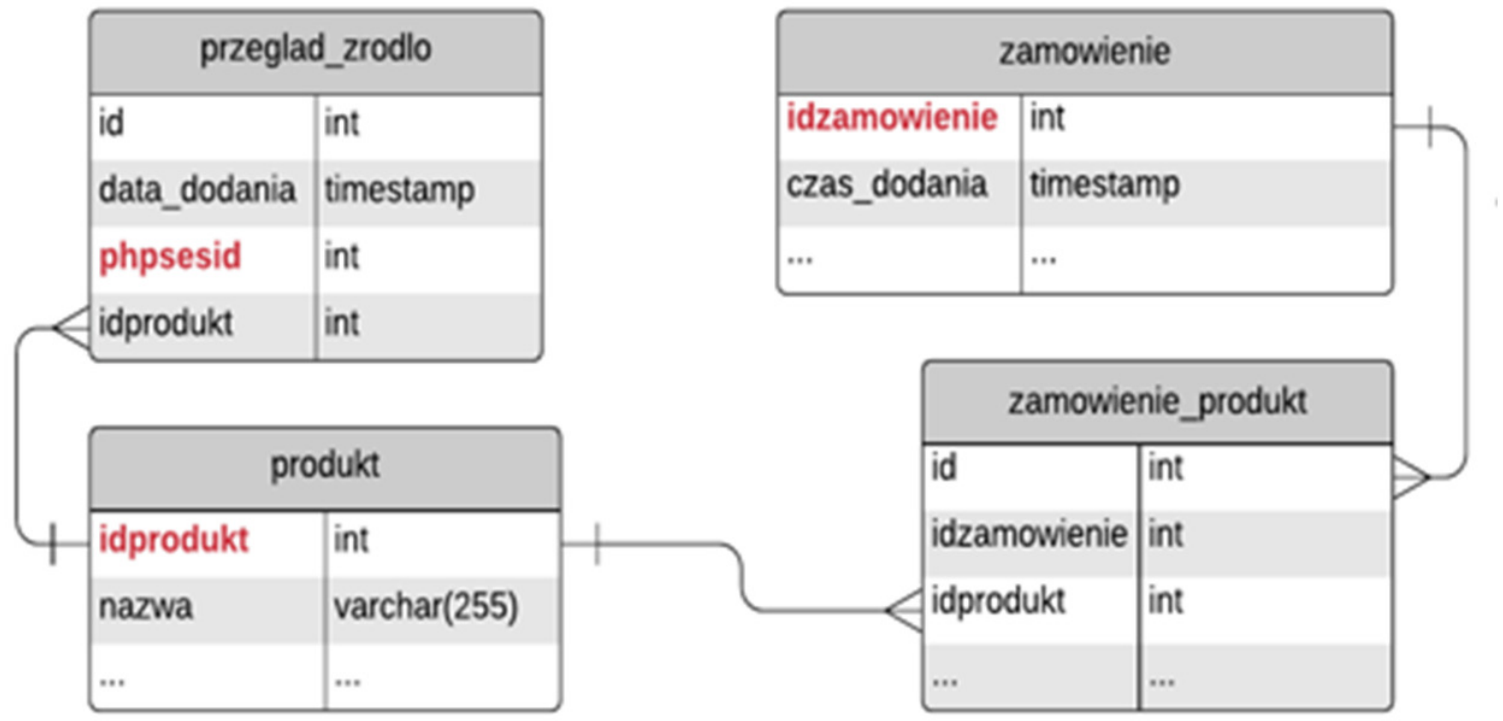

**Fig. 10: Data structure in the form of an ERD diagram of the database of the IT system AM76.pl**

The data used to build graph G was information such as:

$O$ – a collection of objects being products with unique identifiers: idproduct (*produkt.id*)

$J = K_1 \cup K_2$ – collection of kernels made up of two classes:

$K_1$ - visiting a shop with unique identifiers: phpsesid (view_source.phpsesid)

$K_2$ - the purchase orders with unique identifiers: purchaseorderid (*purchaseorder.id*)

$E = E_1 \cup E_2$ – a collection of edges composed of:

$E_1$ - visiting the products: phpsesid –> idprodukt

$E_2$ - the items of the purchase orders: purchaseorderid –> productid

$K_1$ — $J$

| phpsesid | data |
|---|---|
| 1624844680 | 2019-12-15 |
| 1601766927 | 2019-12-15 |
| 1372428021 | 2019-12-15 |
| 1630107088 | 2019-12-15 |
| 1539705806 | 2019-12-15 |

$K_2$ — $J$

| idzamowienie | data |
|---|---|
| 89031 | 2019-12-15 |
| 89032 | 2019-12-15 |
| 89033 | 2019-12-15 |
| 89037 | 2019-12-15 |
| 89048 | 2019-12-15 |

$K_1$ — $E$

| phpsesid | idprodukt |
|---|---|
| 1000180571 | 3454 |
| 1000180571 | 3492 |
| 1000852593 | 4314 |
| 1000852593 | 4050 |
| 1000896693 | 3046 |

$K_2$ — $E$

| idzamowienie | idprodukt |
|---|---|
| 89031 | 4750 |
| 89031 | 4503 |
| 89031 | 5496 |
| 89031 | 4946 |
| 89032 | 3689 |

$O$

| idprodukt | nazwa |
|---|---|
| 2 | Abalone Classic |
| 3 | Abalone Travel (edycja polska) |
| 5 | Agricola |
| 7 | Agricola: Torfowisko |
| 8 | Alhambra (edycja polska) |

**Fig. 11: Examples of data for individual elements of the graph G**

The data described above have been exported to the graphs' database Neo4j. This database allowed to present the data contained in the MySQL relational database in the form of the Graph **G**.

OpenSource software called Cytoscape was used for the research. Cytoscape is a software environment for integrated models of biomolecular interaction networks. The basic version of the above mentioned software has been enriched with a plugin called CypherQueries, which allows for direct connection to the Neo4j database. Neo4j is a native graph database, built from the ground up to leverage not only data but also data relationships.

Within the research, individual steps (K01–K06) of the "ARS" algorithm were executed.

*K01*

The incoming graph **G** containing:

- 1733 nodes (red) in the collection $O$;
- 2056 nodes (blue) in the collection $K_1$;
- 314 nodes (green) in the collection $K_2$;
- 7459 edges in the collection $E_1$;
- 964 edges in the collection $E_2$.

Moreover, the object **m** was an object (node) with the identifier (idproduct) 4537 having the name "Teotihuacan: City of the Gods".

K02

The sub-graph **G'm** of the graph **G** has been constructed. The sub-graph consisted of the node **m** and nodes adjacent to it and edges linking them. The adjacent nodes were the kernels of the sessions, in which the node **m** occurred.

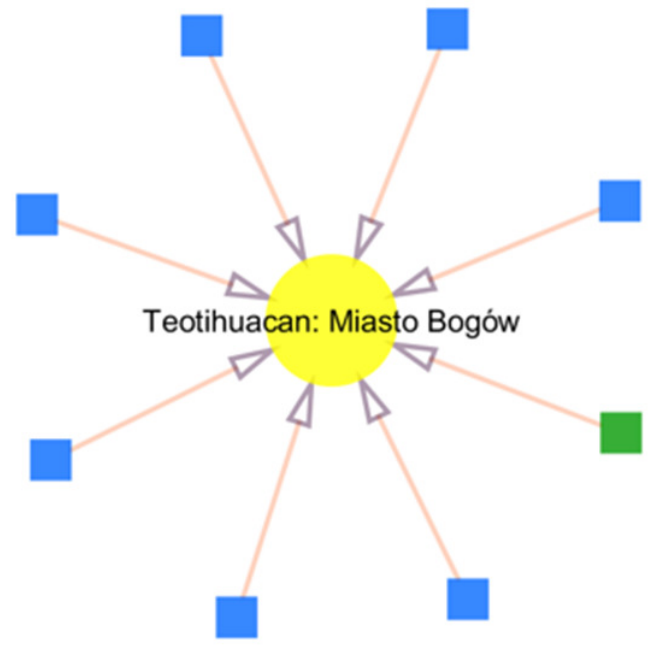

**Fig. 12: The constructed sub-graph G'$_m$ of the graph G**

*K03*

The sub-graph **G''m** of the graph **G** was built. Nodes that are objects related to the kernels of the sessions being in the graph have been added to the subgraph **G'm**.

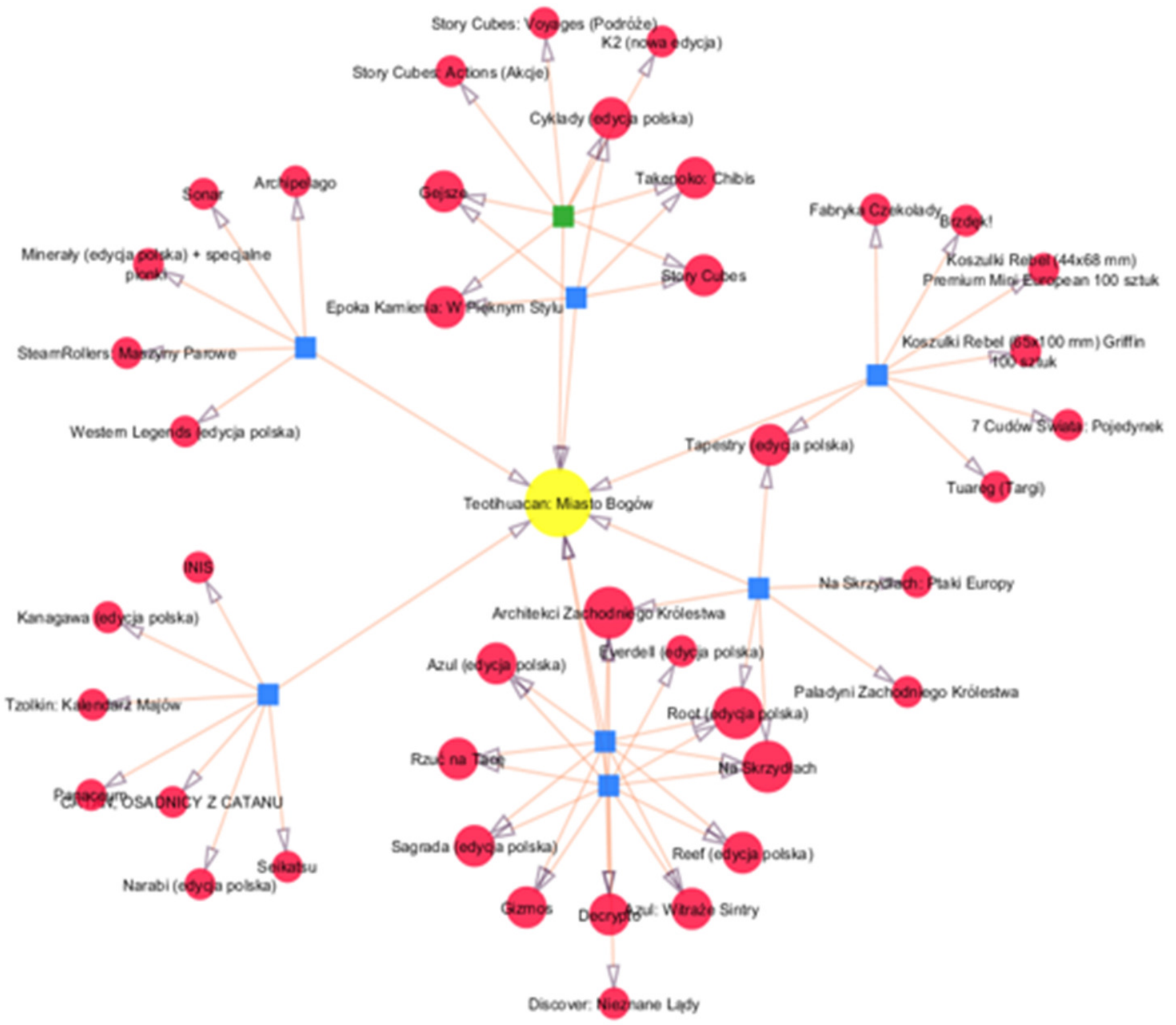

**Fig. 13: The constructed sub-graph G''m of the graph G**

*K04 and K05*

Using the Cytoscape software's analytical module, the incoming steps for each node of the sub-graph **G'm** were estimated, and then the steps were sorted in descending order.

Table Panel

| Indegree | idprodukt | nazwa | _neo4j_labels |
|---|---|---|---|
| 8 | 4537 | Teotihuacan: Miasto Bogów | [obiekt_O] |
| 3 | 4503 | Na Skrzydłach | [obiekt_O] |
| 3 | 4879 | Root (edycja polska) | [obiekt_O] |
| 3 | 4554 | Architekci Zachodniego K... | [obiekt_O] |
| 2 | 4043 | Decrypto | [obiekt_O] |
| 2 | 196 | Story Cubes | [obiekt_O] |
| 2 | 4050 | Azul (edycja polska) | [obiekt_O] |
| 2 | 5496 | Epoka Kamienia: W Piękn... | [obiekt_O] |
| 2 | 2561 | Takenoko: Chibis | [obiekt_O] |
| 2 | 3572 | Gejsze | [obiekt_O] |
| 2 | 4590 | Azul: Witraże Sintry | [obiekt_O] |
| 2 | 4101 | Sagrada (edycja polska) | [obiekt_O] |
| 2 | 4789 | Reef (edycja polska) | [obiekt_O] |
| 2 | 263 | Cyklady (edycja polska) | [obiekt_O] |
| 2 | 4393 | Gizmos | [obiekt_O] |
| 2 | 5448 | Rzuć na Tacę | [obiekt_O] |
| 2 | 5443 | Tapestry (edycja polska) | [obiekt_O] |
| 1 | 784 | Tzolkin: Kalendarz Majów | [obiekt_O] |
| 1 | 4784 | Fabryka Czekolady | [obiekt_O] |
| 1 | 305 | CATAN, OSADNICY Z CA... | [obiekt_O] |
| 1 | 4117 | INIS | [obiekt_O] |
| 1 | 4555 | Paladyni Zachodniego Kr... | [obiekt_O] |
| 1 | 5486 | Minerały (edycja polska) ... | [obiekt_O] |
| 1 | 4729 | Seikatsu | [obiekt_O] |
| 1 | 4098 | SteamRollers: Maszyny P... | [obiekt_O] |

**Fig. 14: The result of sorting the incoming steps of the graph G''m**

*K06*

Eventually, the result of the sorting was saved in the CSV file, omitting the m node with the identifier (idproduct) 4537. This file was a physical representation of the resultant vector **Rm.**

## Evaluation of the algorithm

In order to assess the effectiveness of the algorithm against other recommendation algorithms it is proposed to:

- implement the tested n-algorithms in a real e-commerce system;
- present the results of recommendations to clients based on a random algorithm with a 1/n probability (where n is the number of tested algorithms);
- register the actions taken by users as a collection $D = \{d_1, d_2, d_3, d_4, d_5, d_6, \ldots\}$ in which the action in the n-th step would be represented by the vector
  $d_n = < d_n(1), d_n(2), d_n(3), d_n(4) >$
  where the vector positions mean:
  $d_n(1)$ – the recommendation algorithm used
  $d_n(2)$ – the object, for which the recommendation was made
  $d_n(3)$ – the resulting vector of the recommended objects;
  $d_n(4)$ – the object selected by the client

On the basis of the collected results, a measure of effectiveness for the tested algorithms should be calculated in the form of

$$\boldsymbol{E}(\boldsymbol{a}') = \frac{\sum_{n=1}^{|D|}[d_n(1)=a' \; AND \; (4)\in d_n(3)]}{\sum_{n=1}^{|D|}[d_n(1)=a']} \qquad (5)$$

where:

$a'$ – algorithm for which efficiency is calculated;

$[i]$ – Inverson's operator.

$$[i] = \begin{cases} 1 & gdy\ i\ jest\ prawdziwe \\ 0 & gdy\ i\ jest\ fałszywe \end{cases}$$

$d_n(i)$ – returns the value of vector $d_n$ at the position $n = 1 \div 4$

## Modifications and development

The base form of the Recommendations Session Algorithm can be developed in various directions, among other such as:

Adding a third layer of session nodes

It consists in adding consecutive layer of nodes (kernels and objects), creating the sub-graph $\mathbf{G''_m}$, adjacent to the nodes of the sub-graph $\mathbf{G''_m}$.

In addition, further algorithm steps such as:

- calculating the outgoing steps of all nodes;
- deletion of the step 1 nodes (sessions with one object);
- calculating the incoming steps of nodes;

In this way, the resulting vector of recommendation $\mathbf{R_m}$, for object **m,** shall have higher values of parameters indicating the position of the recommendation, i.e. the incoming steps, without the need to expand graph **G**.

The additional cost that will then be incurred will be linked to additional calculations.

Adding weights to edges

At the stage of preparation of the graph **G**, the weights' values should be related to the individual classes of the session's kernels. Then assign these weights to the edges coming out of them. As a result, the incoming steps of objects shall be the sum of the weights of the incoming edges.

Thanks to this procedure the property shall be given that some session classes, such as expert recommendations, are more valuable than random visits by an anonymous user.

Pathway of a user

In order to eliminate the algorithm's historic agnosticism, an auxiliary variable should be implemented in the form of the vector **M**, in which information about "viewed" objects shall be stored. The first element of this table would be the node **m**, which after a change would be called $\mathbf{m_1}$. For the subsequent elements of the table **M**, indexes would be the number of steps in which the user would "visit" subsequent nodes. The steps of the algorithm would then depend not on a single node **m**, but on the vector of nodes **M**.

Pathway of recommendation

As an impulse for the development of the next path of the algorithm, one may consider the willingness to receive as a result not only a proposal for the next step (recommendation vector $\mathbf{R_m}$), but the whole path of recommended products ending with a satisfactory state, for example, a purchase. For this purpose, the theory related to the Semi-markowski processes of the CD class (continuous over time, discrete in their states) can be used. In this case, the modifications would be much larger and, above all, would involve different conversion of the source data and construction of the Graph **G**.

## Conclusions

The presented algorithm that is the member of the group of recommendation algorithms is relatively simple. The biggest difficulty is to prepare the input data for this algorithm in the form of Graph **G**. This difficulty is especially visible in the aspect of extraction and export of real data from functioning systems (mostly relational databases) to the graph-network structure (Graph **G**).

The algorithm, in its basic form, is historically agnostic in relation to users' actions. For the algorithm, the current history of user actions is irrelevant.

Furthermore, the characteristics of the users of the algorithm are irrelevant for the algorithm. This may be both a disadvantage (lack of personalization) and an advantage (meeting the GDPR requirements). The algorithm gives the same results in each call for various users. The only variables affecting the algorithm are the selected object for which a recommendation is made and the Graph **G**.

The practical usefulness of the results of the algorithm may give rise to discussions due to the very similar values of the parameter indicating the position of the recommendation, i.e. the incoming steps. However, this feature is very much dependent on the size of the Graph **G**, and precisely on the number of edges in the graph. The bigger is the number of the edges, the higher the degree of nodes in the graph.